\documentclass{ws-procs975x65}
\usepackage{graphicx}

\newcommand{\Neff}{\ensuremath{N_{\rm eff}}}
\newcommand{\mnu}{\ensuremath{\Sigma m_{\nu}}}
\newcommand{\lcdm}{$\Lambda$CDM}
\newcommand{\hou}{\ensuremath{\mbox{ km s}^{-1}\mbox{Mpc}^{-1}}}

\begin{document}

\title{Neutrino Properties and the Cosmological Tensions in the $\Lambda$CDM Model}
\author{Stefano Gariazzo}

\address{Instituto de F\'isica Corpuscular (CSIC-Universitat de Val\`encia),\\
Paterna (Valencia), Spain\\
E-mail: gariazzo@ific.uv.es}

\begin{abstract}
In this talk I will review the current status of the constraints on the neutrino properties
from cosmological measurements, with a particular focus on their mass and effective number.
I will also discuss the existing tensions within the context of the \lcdm\ model,
including the discrepancies on the Hubble parameter and on the matter fluctuations at small scales,
and how neutrinos could help to alleviate the aforementioned problems.
\end{abstract}


\bodymatter

One of the most amazing experimental results of the recent years has been achieved by the Planck collaboration,
which measured the Cosmic Microwave Background (CMB) radiation
with an unprecedented precision \cite{Adam:2015rua,Ade:2015xua,Akrami:2018vks,Aghanim:2018eyx}.
Such result allowed us to improve our knowledge of the Universe history,
which can be very well described by the six-parameters model named \lcdm,
after the cosmological constant ($\Lambda$) and the Cold Dark Matter (CDM) fluids,
which represent most of the current energy density in the Universe.

Within the context of the \lcdm\ model and using CMB data,
we can derive constraints on a number of cosmological quantities,
as for example the Hubble parameter $H_0$ or the matter perturbations at small scales,
usually quantified through the parameter $\sigma_8$,
which describes the mean matter fluctuations in a sphere with a radious of $8h^{-1}$~Mpc.
The values of $H_0$ and $\sigma_8$ obtained from a fit of the CMB data within the context of the \lcdm\ model
are not in agreement with the values that are measured in the local universe, at small redshifts.
For instance, while Planck data \cite{Ade:2015xua} point towards $H_0=67.27\pm0.66$\hou,
estimates from the local Universe \cite{Riess:2016jrr} indicate $H_0=73.24\pm1.74$\hou:
a $\sim3.4\sigma$ tension exists.
In the same way, cosmological and local determinations of $\sigma_8$ typically
exhibit a $\sim2\sigma$ tension with local determinations,
as performed for example by the KiDS \cite{Joudaki:2016kym} or DES \cite{Abbott:2017wau} experiments.
These tensions may be the result of an incomplete understanding of systematics in the experimental determinations
or of the presence of new physics which is not considered in the \lcdm\ model.

Under a different point of view,
cosmology is nowadays one of the most powerful tools to study particular aspects of particle physics,
such as some neutrino properties, and in particular their absolute masses
or the way the interact and thermalize in the early universe.
In the standard cosmological analyses, neutrinos are usually treated in the minimal way,
assuming minimal masses ($\mnu=0.06$~eV, corresponding to a massless lightest neutrino and two massive neutrinos,
for which the masses come from the mass splittings measured in neutrino oscillation experiments
and assuming normal ordering, see e.g.\ Ref.~\refcite{deSalas:2017kay})
and a contribution to the effective number of relativistic species
in the early Universe $\Neff=3.05$ \cite{deSalas:2016ztq}.
Non-standard neutrino properties may alter the cosmological evolution
and have an effect on the CMB spectrum, as well as the formation of structures at later times,
so that both parameters (\mnu\ and \Neff) can be well constrained.

First of all, neutrinos decoupled in the early universe when they were still relativistic and contribute to \Neff.
Variations of \Neff\ influence the universe evolution before the photon decoupling,
and as a consequence the CMB spectrum through a variation of the amplitude of the peaks and their angular scale.
In order to compensate the effects of having \Neff\ different from 3.05, as it would be in non-standard cases
(modified neutrino properties or presence of new neutrino-like particles),
one can alter the energy densities of CDM and of dark energy.
In this case, the angular scale of the CMB peaks can be left unvaried, at the expense of increasing $H_0$
and having a damping of the CMB spectrum at small scales.

After the non-relativistic transition,
which occurs at different times for each neutrino eigenstate depending on its mass,
neutrinos contribute as regular massive particles and must be considered
when computing the energy density of matter.
Their influence on the CMB spectrum is related to the modification of the late-time expansion,
both through the variations of the amplitude of the first peak,
related to the early integrated Sachs-Wolf (ISW) effect,
and of the angular position of the CMB peaks,
which comes from the modifications of the angular diameter distance in presence of massive neutrinos.
This latter effect can be partially compensated with a rescaling of the Hubble parameter $H_0$,
so that an anti-correlation between $H_0$ and \mnu\ appears.

Massive neutrinos, however, have a proportionally more important impact
on the evolution of large scale structures than on the CMB spectrum.
Neutrinos, indeed, possess a large thermal velocity which allows them
to escape the gravitational attraction at very small scales.
For this reason, no neutrino overdensity can grow at scales smaller than a characteristic scale,
named the free-streaming scale, which depends on the neutrino mass.
When one compares two universes which share the same total amount of matter,
one with massless and the other with massive neutrinos,
the universe with massive neutrinos will therefore have a suppressed matter power spectrum at the smallest scales,
because the free-streaming impedes the growth of neutrino perturbations and meanwhile
reduces the possibility for the gravitational attraction to accrete the total matter overdensities.
The fact that for massive neutrinos there is a reduction of the matter fluctuations at small scales
means that an increase of the neutrino masses correlates with a decrease of $\sigma_8$.

Given these main effects that neutrinos produce on the cosmological quantities and
the resulting degeneracies between the neutrino properties (\mnu\ and \Neff), $H_0$ and $\sigma_8$,
one may be tempted to say that neutrinos can reduce or even solve
the tensions that appear within the context of the \lcdm\ model between the CMB preferred cosmology
and local universe probes.
Unfortunately, this does not happen.
The analyses of CMB data, indeed, show that it is difficult to alleviate both the tensions at the same time.
If one considers CMB measurements from Planck 2015 \cite{Ade:2015xua},
the best fit values for \Neff\ and \mnu\ lie very close to the expected values:
$\Neff=3.0\pm0.2$ (68\% CL) and $\mnu<0.17$~eV (95\% CL).
Naively, in order to allieviate the $H_0$ tension, an higher \Neff\ would be useful.
In the same way, heavier neutrinos would allow to reduce the matter fluctuations at small scales, $\sigma_8$.
The problem is that an increase in \Neff\ requires to be compensated by a larger matter density at all times,
so that at the practical level also $\sigma_8$ increases.
In the same way, heavier neutrinos require a smaller $H_0$:
it is therefore easy to see that the two tensions cannot be solved at the same time
solely varying neutrino properties.

One last word on Planck measurements in connection with neutrino masses.
The Planck 2015 estimates of CMB lensing from the distortions of the TT spectrum \cite{Ade:2015xua}
are in slight tension with the reconstruction of the lensing profile
from the 4-point correlation function \cite{Ade:2015zua},
as quantified by the phenomenological lensing amplitude parameter, $A_L=1.22\pm0.10$ (68\% CL),
which is expected to be 1 in general relativity.
Massive neutrinos, again thanks to their free-streaming, can allieviate the tension.
This is the reason for which the 95\% CL limits change
from $\mnu<0.17$~eV (Planck TT,TE,EE + lowP + BAO)
to $\mnu<0.22$~eV (Planck TT,TE,EE + lowP + BAO + lensing)
when the lensing information is considered in the analyses.

Let us now discuss non-standard neutrino degrees of freedom.
In the very recent years, for the first time we have had model-independent indications that neutrino oscillations
at a baseline that is not compatible with the standard three-neutrino oscillations may exist:
the NEOS~\cite{Ko:2016owz} and DANSS~\cite{Alekseev:2018efk}
reactor experiments observed electron antineutrino disappearance
at distances of $\sim24$~m and $\sim10-12$~m, respectively.
Combined model-independent analyses \cite{Gariazzo:2018mwd,Dentler:2017tkw} show
that the combination of the two experiments
give a $\sim3.5\sigma$ preference for the three active plus one light sterile (3+1) neutrino scenario
over the standard three neutrino case,
implying the existence of a fourth neutrino, with a mass around 1.1~eV.
When considering the global picture,
one must also take into account data from muon neutrino disappearance
(e.g.\ from MINOS+ \cite{Adamson:2017uda} or IceCube \cite{Aartsen:2017bap,TheIceCube:2016oqi})
or electron neutrino appearance in a muon neutrino beam
(e.g.\ from LSND \cite{Aguilar:2001ty} or MiniBooNE \cite{Aguilar-Arevalo:2018gpe}).
The first problem is that there is a tension between appearance (lead by LSND and MiniBooNE)
and disappearance data (lead by NEOS, DANSS and MINOS+),
see e.g.\ Ref.~\refcite{Dentler:2018sju},
so that the situation is still unclear.
The second problem is that if the 3+1 neutrino oscillation parameters as obtained from the global fit
are considered to compute the neutrino decoupling in the early universe,
one would have a fully thermalized fourth neutrino, with $\Neff\simeq4$ \cite{Hannestad:2012ky,Mirizzi:2012we},
which is in clear tension with the constraints on \Neff.
This means that if the light sterile neutrino will be confirmed,
some additional mechanism will be needed to suppress its thermalization in the early universe.
As in the case of active neutrinos,
the additional sterile neutrino and its properties can only marginally reduce the tension between
the local universe probes of $H_0$ and $\sigma_8$ and the corresponding values inferred by CMB observations
in the context of an extended \lcdm\ model.

As a summary,
our current knowledge of the universe and its history shows a number of mild tensions,
in particular regarding the Hubble parameter $H_0$ and
the matter fluctuations at small scales, encoded by $\sigma_8$.
A mild tension is also present among the two different determinations of the CMB lensing by Planck \cite{Ade:2015xua}.
Neutrino properties (considering both active or sterile states)
can alleviate each of these tensions if considered separately,
but in the global picture the situation cannot be easily solved.
If these cosmological tensions or the existence of a light sterile neutrino will be confirmed in the future,
some new physical mechanism will be needed in order to obtain a consistent model that can explain
the entire evolution of the universe.

\section*{Acknowledgments}
Work supported by the European Union's Horizon 2020 research and innovation programme
under the Marie Sk{\l}odowska-Curie individual grant agreement No.\ 796941,
and by the Spanish grants SEV-2014-0398 and FPA2017-85216-P (AEI/FEDER, UE),
and PROMETEOII/2018/165 (Generalitat Valenciana).

\bibliography{biblio}

\end{document}